\documentclass[runningheads]{llncs}
\usepackage[T1]{fontenc}
\usepackage{graphicx}
\usepackage{amsmath}
\usepackage{multirow} 
\usepackage{pdflscape} 
\usepackage{hyperref} 
\usepackage{xcolor} 
\usepackage{subcaption} 
\usepackage{amssymb} 
\usepackage{pifont}
\usepackage{tcolorbox}
\usepackage{xspace}
\usepackage{array} 
\usepackage{tabularx}
\usepackage{orcidlink}

\usepackage{colortbl} 

\newcolumntype{Y}{>{\centering\arraybackslash}X}

\hypersetup{
    colorlinks=true,
    linkcolor=blue,
    filecolor=magenta,      
    urlcolor=blue,
    pdftitle={Pricings Report},
    pdfpagemode=FullScreen,
}

\usepackage{ulem} 
\newcommand{\chunk}[2]{}

\newcommand{\agarcia}[1]{
\chunk{Alejandro}{\textbf{\textcolor{orange}{\textsl{#1}}}}
}

\newcommand{\aruiz}[1]{
\chunk{Antonio}{\textbf{\textcolor{blue}{\textsl{#1}}}}
}

\newcounter{rqAnswerCounter}
 
\newenvironment{rqAnswer}
  {\begin{list}
    {\arabic{rqAnswerCounter}}
    {\usecounter{rqAnswerCounter}
     \setlength{\labelwidth}{1em}
     \setlength{\labelsep}{0.5em}
     \setlength{\itemsep}{0pt}
     \setlength{\leftmargin}{1.5em}
     \setlength{\rightmargin}{0em}
     \setlength{\itemindent}{0em}
     
    }
  }
{\end{list}}

\newcommand{\powerset}[1]{\mathbb{P}(#1)}

\begin{document}

%
\title{Racing the Market: An Industry Support Analysis for Pricing-Driven DevOps in SaaS}

%
\titlerunning{An Industry Support Analysis for Pricing-Driven DevOps in SaaS}

\author{Alejandro García-Fernández\inst{1}\orcidlink{0009-0000-0353-8891} \and
José Antonio Parejo\inst{1}\orcidlink{0000-0002-4708-4606} \and
Francisco Javier Cavero\inst{1}\orcidlink{0009-0004-2453-8814} \and
Antonio Ruiz-Cortés\inst{1}\orcidlink{0000-0001-9827-1834}}
\authorrunning{A. García-Fernández et al.}
%
\institute{SCORE Lab, I3US Institute, Universidad de Sevilla, Spain \\
\email{\{agarcia29,japarejo,fcavero,aruiz\}@us.es}}
\maketitle              
\begin{abstract}

The SaaS paradigm has popularized the usage of pricings, allowing providers to offer customers a wide range of subscription possibilities. This creates a vast configuration space for users, enabling them to choose the features and support guarantees that best suit their needs.
Regardless of the reasons why changes in these pricings are made, the frequency of changes within the elements of pricings continues to increase.
Therefore, for those responsible for the development and operation of SaaS, it would be ideal to minimize the time required to transfer changes in SaaS pricing to the software and underlying infrastructure, without compromising the quality and reliability.

This work explores the support offered by the industry for this need. By modeling over 150 pricings from 30 different SaaS over six years, we reveal that the configuration space grows exponentially with the number of add-ons and linearly with the number of plans. We also evaluate 21 different feature toggling solutions, finding that feature toggling, particularly permission toggles, is a promising technique for enabling rapid adaptation to pricing changes. Our results suggest that developing automated solutions with minimal human intervention could effectively reduce the time-to-market for SaaS updates driven by pricing changes, especially with the adoption of a standard for serializing pricings.
\end{abstract}

\section{Introduction}
The Software as a Service (SaaS) model is a distribution and licensing paradigm that has grown significantly in popularity over recent decades \cite{Jiang2009}. This model involves the delivery of software through the cloud, providing users with a set of features and support guarantees accessible by paying a periodic fee.

Information about the available fees is provided through a pricing, a structure consisting of various plans and optional add-ons that group and control access to features, imposing usage limits when needed. With this approach, SaaS providers can offer different sets of features and usage limits for different customer profiles, adjusting to varying budgets and requirements. In this way, pricings enhances flexibility for users, enabling them to suit their subscriptions to their needs and easily upgrade or downgrade their plans, and maximizes revenue for providers by catering to varying customer needs and promoting market expansion; as a variety of pricing plans attract a broader customer base.


As this paper demonstrates, changes in pricings are frequent (removal, modification, or addition of features, usage limits, plans, or add-ons). Therefore, to remain competitive, SaaS developers and operators must minimize the time needed to implement these pricing changes in the software and underlying infrastructure (typically a Platform as a Service), while maintaining quality and reliability. This process, which imposes significant challenges on development and operations teams and should be streamlined for efficiency, has been recently coined as \emph{Pricing-driven Development and Operation of SaaS} \cite{JCIS2024LONG}\footnote{For brevity, we may also refer to this concept as Pricing-driven SaaS DevOps.}.

Our goal is to pave the way for further research in the design and development of technologies that automate and optimize the Pricing-driven SaaS DevOps process. 
In order to obtain a deeper and detailed knowledge about the current support offered by the industry for Pricing-driven SaaS DevOps, in this paper we explore the dimensions of change in SaaS pricing by modeling more than 150 pricing plans from 30 different commercial SaaS, and tracking their evolution over six years (2019-2024). In addition, since feature toggling \cite{FOWLER2023} 
allows specific features to be enabled or disabled without deploying new code ---merely by modifying configuration files--- and it is one natural approach for implementing pricings in SaaS source code, 
this paper analyzes its current support for Pricing-driven SaaS DevOps in the industry, academy and open source community. In particular, we focus on feature toggling tools, examining up to 21 in detail.

This work includes the following original contributions in the context of the Pricing of SaaS: 
\begin{enumerate}
    \item The definition of the configuration and evaluation spaces of a pricing, two key concepts for analysing the evolution of pricings. 
    \item An unprecedented analysis of 162 pricings of 30 commercial SaaS, yielding significant insights into the structures and trends of SaaS pricing models.
    \item A comprehensive dataset and repository of these 162 pricing models, derived from the 30 commercial SaaS analyzed \cite{LABPACK}, by using the only metamodel proposed in the literature---Pricing4SaaS \cite{CAISEFORUM24}---, which will facilitate further analysis and research on SaaS pricing.
    \item A comparative overview of the support offered by 21 feature-toggling-based solutions for implementing changes in pricings.
\end{enumerate}

The remainder of this paper is structured as follows: Section \ref{sec:background} introduces SaaS pricings and feature toggles. Section \ref{sec:PricingStructuresObservationalStudy} presents our observational studies, outlining the research questions, the methodology and the results obtained. Section \ref{sec:threadsToValidity} outlines threats to validity. Section \ref{sec:relatedWork} describes related work. Finally, Section \ref{sec:conclusions} draws conclusions and discusses future lines of research.

\section{Background and Motivation}
\label{sec:background}



\subsection{SaaS pricing}
\label{sec:background:saasPricing}

\begin{figure}[htb]
    \centering
    \includegraphics[width=\textwidth]{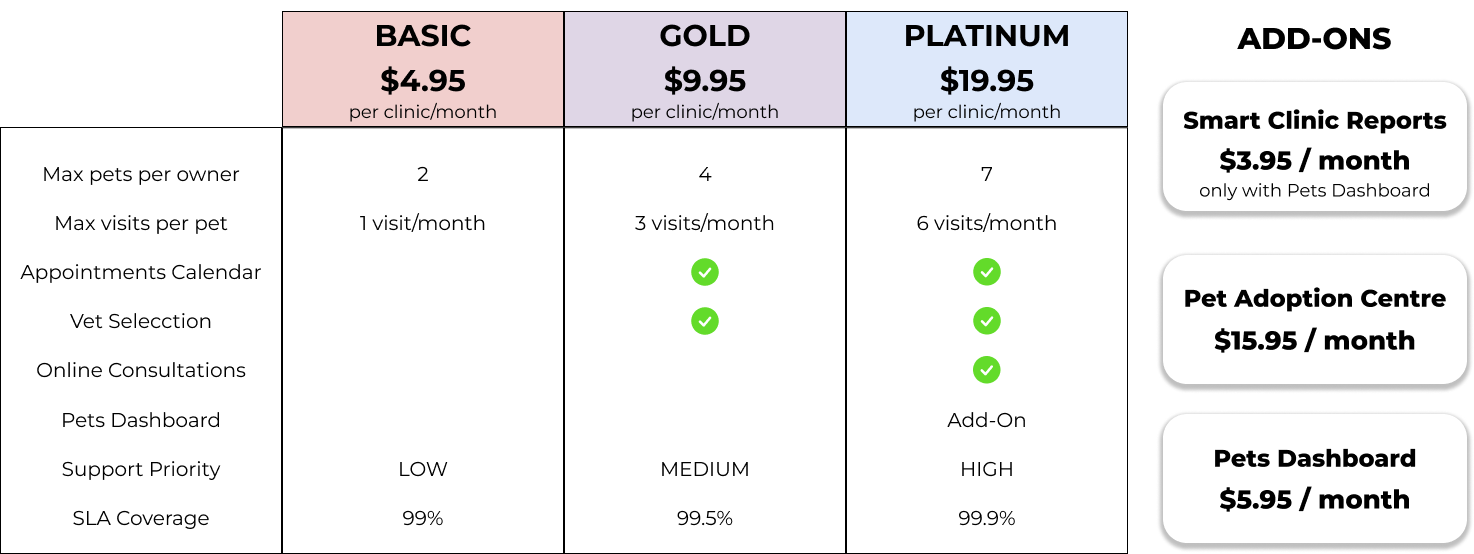}
    \caption{Sample pricing with ten features, three plans and three add-ons. 
    }
    \label{fig:petclinicPricing}
\end{figure}

A pricing is a structure that organizes the \emph{features} of a service ---defined as the distinctive characteristics whose presence/absence may guide an user’s decision towards a particular subscription \cite{JCIS2024LONG}--- into \emph{plans} and \emph{add-ons} to control users access to such features. 
While users can only subscribe to one of the available plans, they can subscribe to as many add-ons as they want since they are available for the contracted plan. 
Fig. \ref{fig:petclinicPricing} illustrates a pricing for PetClinic, a sample veterinary clinic management service that developers use to illustrate the features of a particular software framework or technology in a real-world scenario.\footnote{The base version of PetClinic for Spring can be found \href{https://github.com/spring-projects/spring-petclinic}{here}.} It includes ten features regulated by three plans and three add-ons (seven by plans, three by add-ons), with one add-on exclusive to the PLATINUM plan, and imposes \textit{usage limits} on the ``pets'' and ``visits'' features. Given this running example, it is important to note that not all pricing features are necessarily translated into code within the service. Those that are will be referred to as \textit{functional features}, while those that aren't will be referred to as \textit{non-functional features}. The latter represent service-level guarantees, such as ``Support Priority'' and ``SLA Coverage'' in PetClinic. 

\begin{figure}[htb]
    \centering
    \includegraphics[width=0.9\linewidth]{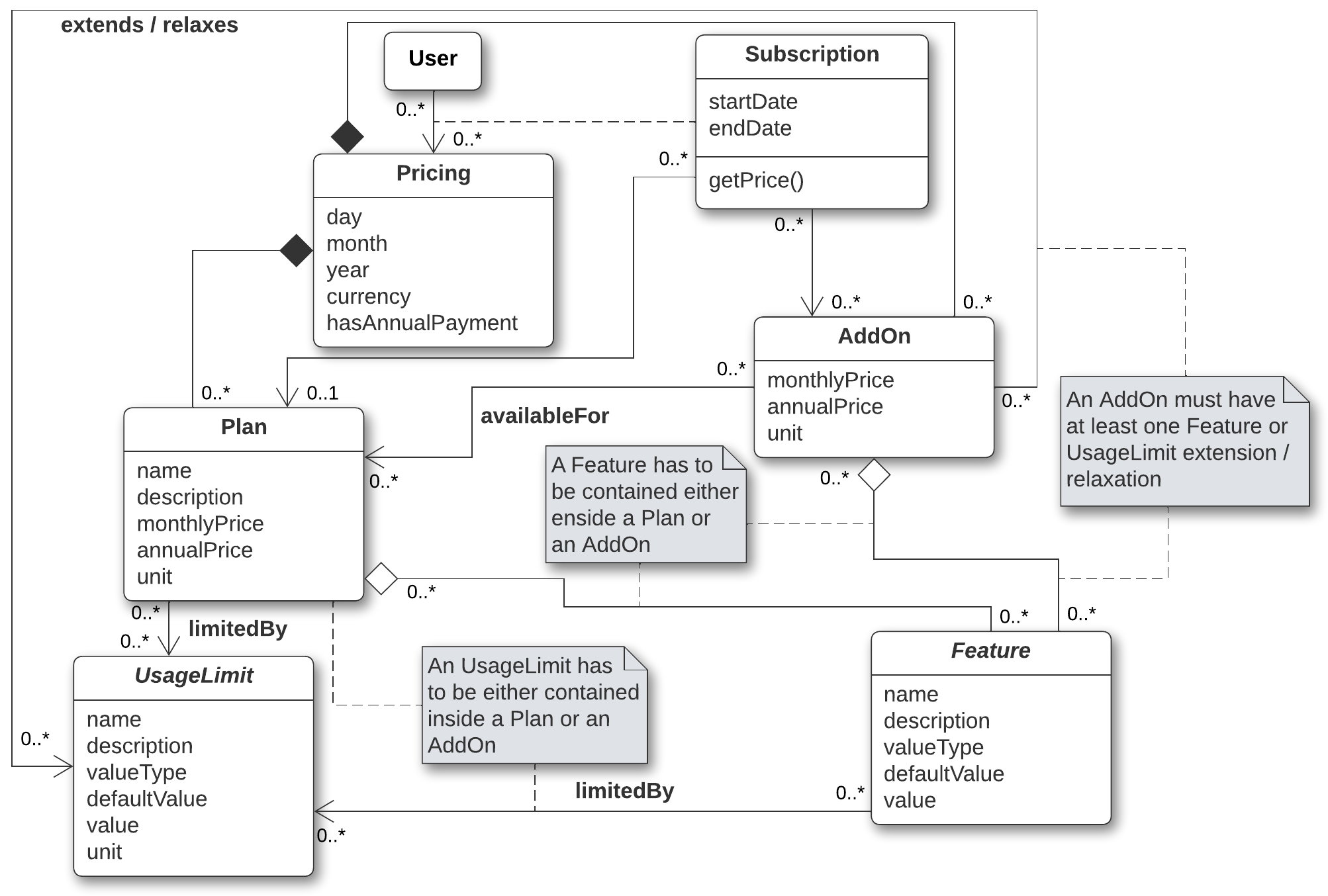}
    \caption{Excerpt of Pricing4SaaS UML Model. Created from the original in \cite{CAISEFORUM24}.}
    \label{fig:pricing4saas}
\end{figure}

According to our research, Pricing4SaaS (see Fig. \ref{fig:pricing4saas}) is the only metamodel proposed in the literature that formalizes these pricing elements \cite{CAISEFORUM24}, being capable of representing pricings regardless of their elements distribution. Additionally, its YAML-based serialization, Yaml4SaaS, has been pivotal in our analysis of industry SaaS (see Section \ref{sec:PricingStructuresObservationalStudy}), allowing us to model all the studied pricings while ensuring portability for our work, making it reusable for future research.



Given a SaaS pricing  with a set of plans $P = \{p_1, p_2, \dots , p_n\}$ and add-ons $A = \{a_1, a_2, \ldots , a_m\}$, a customer interacts with the service by establishing \emph{a subscription}, i.e. a ``bundle'' that may include a plan and optionally a set of add-ons, ensuring that: 

\begin{enumerate}
    \item Subscription is not empty.
    \item It contains exactly 0 plans if \(P = \emptyset\), and 1 otherwise.
    \item Any add-on included within the subscription is not excluded for the selected plan, e.g. the ``Pets Dashboard'' add-on of PetClinic is only available for the PLATINUM plan, meaning the add-on is excluded for BASIC and GOLD plans (\(E(``PetsDashboard'') = \{BASIC,GOLD\}\)).
    \item All required add-ons for any add-on in the subscription are also included in it, e.g. in PetClinic: \(D(``PetsDashboard'') = \{``SmartClinicReport''\}\).
\end{enumerate}


Once customers have made their selection, they commit to paying a periodic fee to gain the ability to access and leverage the features provided by the SaaS in the terms and usage limits set out by the chosen subscription. E.g. customers with the BASIC plan in PetClinic can register up to two pets in their account, cannot select a vet for their visits, etc.

Given this structure, determining the set of different subscriptions within a pricing may become very challenging. We have coined this concept as the \emph{configuration space} of pricings, aiming to enhance our understanding of pricings. Formally, we describe a pricing's configuration space $C$ as:

\begin{center}
    \( C(P,A,E,D) = \{(p,a) \in P \cup \emptyset \times \powerset{A} \mid isValid(p,a,E,D) \)\}
\end{center}

\noindent where $isValid$ encodes the above defined rules and is defined as:
\[
\begin{aligned}
isValid(p, a, E, D) \iff & \left\{ 
    \begin{aligned}
    & p \in P \cup \{\emptyset\}                                && \text{(1)} \\
    & \land a \in \powerset{A}                                   && \text{(1) } \\
    & \land ((p \neq \emptyset) \lor (a \neq \emptyset))           && \text{(1) } \\
    & \land |P| \geq 1 \Rightarrow p \neq \emptyset           && \text{(2) } \\
    & \land \forall a_i \in a, \ p \notin E(a_i)                && \text{(3) } \\
    & \land \forall a_i \in a, \ D(a_i) \subseteq a              && \text{(4)} \\
    \end{aligned} 
\right\}
\end{aligned}
\]

In addition, the maximum cardinality of C is easy to determine, as it is equivalent to not considering the exclusion and dependency relashionships between plans and add-ons ($E = \emptyset$ and $D = \emptyset$). Therefore, the maximum cardinality of a pricing's configuration space with $n$ plans and $m$ add-ons is:

%
%

\[
 max\ |C| = 
 \begin{cases} 
 \ 2^m - 1 & \text{if } P = \emptyset \\ 
 \ n \cdot 2^m & \text{if } P \neq \emptyset
 \end{cases}
 \]

As an illustration, the maximum cardinality of PetClinic's pricing configuration space would be:

\[
max|C| = n \cdot 2^m = 3 \cdot 2^3 = 24
\]

In this regard, add-ons play a crucial role in pricings, since they exponentially increase the size of the configuration space ---enabling the accommodation of a wider range of user needs--- while reducing customer decision fatigue. As the \textit{Paradox of Choice} posits, ``human beings tend to feel less satisfied with their decisions when faced with a greater number of alternatives to choose from'' \cite{schwartz2015paradox}.

\subsection{Feature toggles for Pricing-driven DevOps}
\label{sec:featureToggles}

\begin{figure}[htb]
    \includegraphics[width=\linewidth]{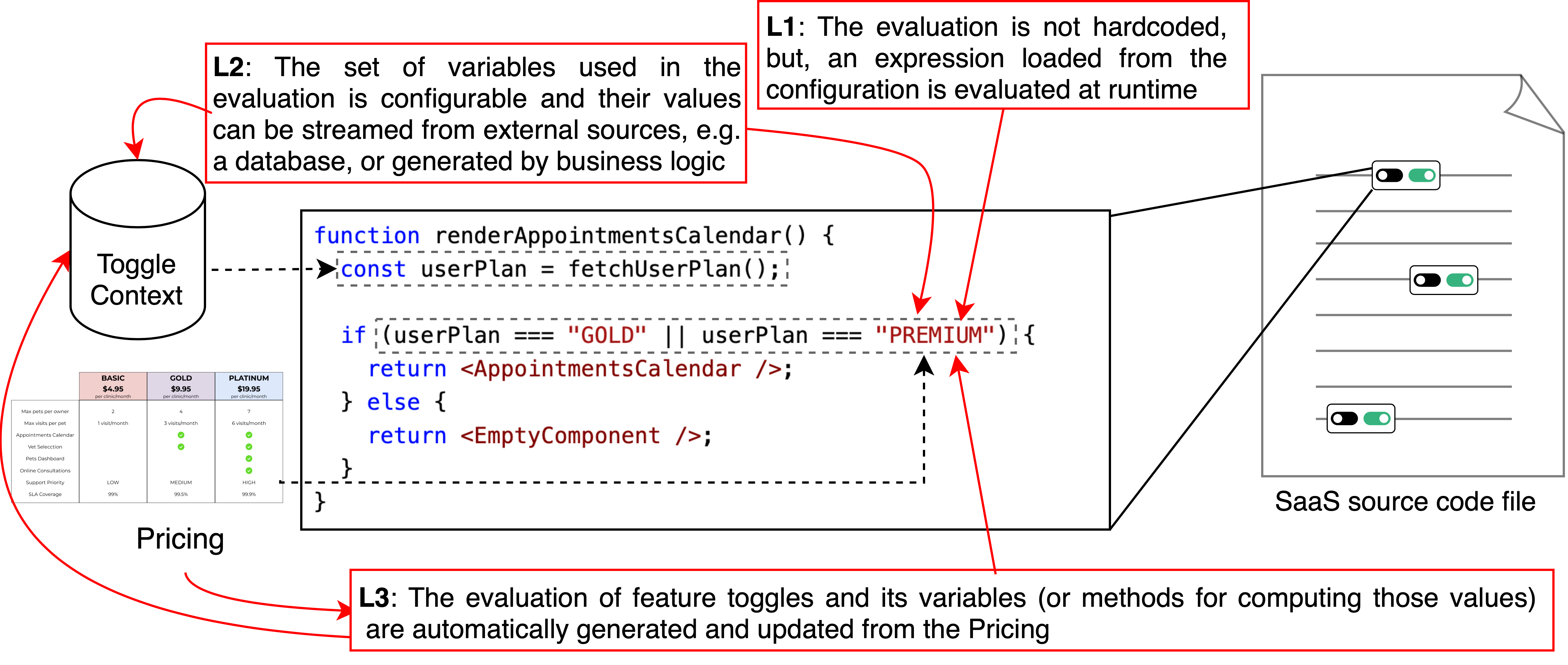}
    \caption{Feature toggles and how the defined capability levels relate to their elements. Based on the original from \cite{JCIS2024LONG}.}
    \label{fig:Feature-toggles-criteria}
\end{figure}

\emph{Feature toggles} are a software development technique that allows features to be dynamically enabled or disabled without modifying the code. In a nutshell, this behaviour is implemented by using boolean expressions on which some values can be assigned, or modified, at runtime, thus providing dynamic evaluations \cite{FOWLER2023}. Fig. \ref{fig:Feature-toggles-criteria} illustrates the simplest version of a feature toggle in the source code of a hypothetical implementation of PetClinic. This toggle evaluate the feature ``Appointments Calendar'' (see Fig. \ref{fig:petclinicPricing}), enabling or disabling its web component based on the user's plan. As shown, the evaluation of the conditional block is hard-coded, but some values depend on dynamic data, e.g. ``userPlan'', which is retrieved, using the ``fetchUserPlan'' function, from the toggle context, an external source (e.g. a database) that contains the data needed for the evaluation.

In this scenario, interpreting a pricing as a software artifact that determines the behavior of the SaaS can be highly effective, as developers can translate it into pricing-driven feature toggles, i.e. permission feature toggles \cite{FOWLER2023} that are used to provide each user with a different version of the service at runtime regarding their subscription. This approach leverages the full potential of SaaS, which is realized when multiple customers with diverse requirements can be accommodated within a single application instance \cite{GHADDAR2012}.


Unfortunately, the use of feature toggles also increases the complexity of managing the service, since they generate ``one of the worst kinds of technical debt'' \cite{TERNAVA2022} and transform testing into a combinatorial problem \cite{Rahman2016}. The concept of pricing's \textit{evaluation space} that we introduce becomes crucial in this context. Given that feature toggles have a inherent evaluation, and considering that any functional feature of the pricing has an associated feature toggle, the evaluation space can be defined as the minimum set of feature toggles that have to be evaluated in order to operate the functional features governed by a pricing, it enables DevOps teams to predict the impact of changes on pricing elements within the service. In order to approximate its size, the simplest architecture must be considered, i.e. monolithic client-server, with each feature evaluated only once on each side. For instance, in PetClinic (Fig. \ref{fig:petclinicPricing}), since only eight features of its pricing are functional (all except ``Support Priority'' and ``SLA Coverage''), the size of the evaluation space of its pricing is 16. Respectively: i) eight pricing-driven feature toggles on the frontend to activate/deactivate UI elements related to the specified pricing features, and ii) the eight corresponding toggles on the backend to manage access to such features. 

As can be seen, the size of the evaluation space usually doubles the number of functional features governed by the pricing, highlighting the challenge of managing pricings with a large feature set. In addition, the number of features is not the only dimension of growth of the evaluation space; although the configuration space doesn't impact on the number of feature toggles required to operate a pricing-driven service, it affects the complexity of the evaluations these toggles perform. Recovering the example from Fig. \ref{fig:Feature-toggles-criteria}, having 3 plans (FREE, GOLD, and PLATINUM) within PetClinic means that deciding whether ``Appointments Calendar'' is available for a user or not requires checking if they are subscribed to either the GOLD or PLATINUM plan. If there were only 2 plans (FREE and GOLD), the check would be simpler, as you would only need to verify if the user is subscribed to the GOLD plan. 




However, the complexity associated with the increasing size of the evaluation space can be mitigated by using feature toggling libraries (see Section \ref{sec:FeatureTogglingStudyMethodology}). The advantage of using such tools to manage pricing-driven feature toggles is that their evaluations don't need to be hard-coded, as they are abstracted into a configuration file. Unfortunately, not all industry feature toggling frameworks provide enough flexibility to efficiently manage pricing-driven feature toggles, so we propose a set of levels based on their capabilities:


\begin{itemize}
    \item \emph{L1: Configuration-based toggling.} Feature toggling tools that meets this level would allow to transfer the evaluation expression of feature toggles to a configuration file that can be modified at runtime to apply changes within the service. 
    However, developers still need to maintain and update the configuration file for each change on the pricing.
    
    \item \emph{L2: 
    Dynamic and extensible contextual evaluation.} In addition to L1, some feature toggling tools allow to customize how the variables used in the toggling expressions, e.g. userPlan in Fig. \ref{fig:Feature-toggles-criteria}, are evaluated at runtime, even loading their values from external sources as needed; so there is no need to declare them before the feature toggle is evaluated. In the running example of Fig. \ref{fig:Feature-toggles-criteria}, this would allow to load the value of ``userPlan'' from an external source, such as the session of the current user.

    \item \emph{L3: Pricing-aware evaluation.}
    This level involves the capability of feature toggling tools to automatically generate dynamic evaluation expressions for feature toggles from a serialized pricing. For instance, the evaluation shown in Fig. \ref{fig:Feature-toggles-criteria} would be replaced by:
    $userSubscription[``appointmentsCalendar"]$
    which dynamically checks the pricing to determine whether the ``Appointments Calendar'' feature is available within the user's subscription. This advanced control allows the tool to automatically adapt to changes in the serialized pricing, ensuring that any updates to the document are seamlessly integrated into the system's functionality and behaviour. 
    
\end{itemize}

\section{Observational studies}
\label{sec:PricingStructuresObservationalStudy}
To meet our goal of paving the way for further research in the design and development of technologies that automate and optimize the Pricing-driven DevOps process for SaaS ---thus providing solutions to previously highlighted issues--- the following research questions were defined:

\emph{$RQ_1$}: \textbf{How are pricings' configuration and evaluation spaces evolving in modern SaaS?} 
Given that in recent years SaaS pricing models have incorporated new elements, such as add-ons, and modified their structure (in terms of plans and features); this research question addresses several key aspects of pricing dynamics to uncover the trends in design and complexity of these structures, thus identifying the challenges faced by SaaS providers to manage Pricing-driven Development and Operation of SaaS.


\emph{$RQ_2$}: \textbf{Are feature toggling tools appropriate for optimizing pricing-driven feature toggling?} Given the dimensions of variability of SaaS pricings, projected into their configuration and evaluation spaces, this research question aims to identify whether a gap exists between the evolution of pricings and the current industry tooling.


\emph{$RQ_{3}$}: \textbf{\textit{Does Yaml4SaaS provide enough expressiveness to represent real world pricings?}} I.e are we able to accurately model all the pricings analyzed in the study for $RQ_1$?

\subsection{Methodology for the analysis of SaaS pricings}
\label{sec:MethodologyForTheAnalysisOfPricingStructures}

\textbf{Sample selection}. As far as we know, there is no systematic methodology for selecting a sample of SaaS pricings. Therefore, to obtain our sample, we decided to start with the 13 SaaS included within the repository of the authors of Pricing4SaaS \cite{CAISEFORUM24}. In our goal to expand their dataset to include a total of 30 SaaS, we extracted some additional services from \cite{saasSource1} with this selection criterion: i) each selected SaaS must have a trackable pricing history with snapshots of their pricing webpage available in the \href{https://web.archive.org}{Wayback Machine} for at least four years between 2019 and 2024, inclusive; ii) their snapshots must contain a clear list of features for most of these years.


\textbf{Snapshot selection}. For each year from 2019 to 2024, priority was given to selecting pricing versions from October/November (when available), aiming to maintain a gap of six months to one year between each studied snapshot. For 2024, the pricing version from June/July was selected, creating a snapshot in the Wayback Machine when needed. Table \ref{table:dataset} presents the resulting dataset of SaaS, detailing the number of snapshots and additional metrics for the latest version of their pricing (2024).

\begin{table}[htbp]
\centering
\begin{tabularx}{\textwidth}{|l|Y|Y|Y|Y|c|l|Y|Y|Y|Y|Y|}
\hline
\textbf{SaaS} &
\textbf{S} & \textbf{F} & \textbf{P} & \textbf{A} & \textbf{C} &
\textbf{SaaS} &
\textbf{S} & \textbf{F} & \textbf{P} & \textbf{A} & \textbf{C} \\ \hline
\rowcolor{gray!20}
\href{https://www.salesforce.com/eu/sales/pricing/}{Salesforce} & 6 & 111 & 3 & 14 & 12544 &
\href{https://buffer.com/pricing}{Buffer} & 6 & 76 & 4 & 3 & 7 \\ \hline
\href{https://github.com/pricing}{GitHub} & 6 & 81 & 3 & 14 & 8960 &
\href{https://www.atlassian.com/software/jira/pricing}{Jira} & 6 & 60 & 4 & 1 & 7 \\ \hline
\rowcolor{gray!20}
\href{https://www.postman.com/pricing/}{Postman} & 5 & 100 & 4 & 12 & 1412 &
\href{https://www.notion.so/pricing}{Notion} & 4 & 58 & 4 & 1 & 7 \\ \hline
\href{https://databox.com/pricing}{Databox} & 6 & 62 & 5 & 8 & 786 &
\href{https://www.figma.com/pricing/}{Figma} & 6 & 90 & 6 & 0 & 6 \\ \hline
\rowcolor{gray!20}
\href{https://www.openphone.com/pricing}{OpenPhone} & 5 & 48 & 3 & 6 & 192 &
\href{https://www.box.com/pricing}{Box} & 6 & 50 & 5 & 0 & 5 \\ \hline
\href{https://www.wrike.com/comparison-table/}{Wrike} & 6 & 78 & 5 & 5 & 85 &
\href{https://www.canva.com/pricing/}{Canva} & 6 & 92 & 4 & 0 & 4 \\ \hline
\rowcolor{gray!20}
\href{https://www.tableau.com/pricing/teams-orgs}{Tableau} & 6 & 41 & 3 & 7 & 48 &
\href{https://www.dropbox.com/plans}{Dropbox} & 4 & 82 & 4 & 0 & 4 \\ \hline
\href{https://zapier.com/pricing}{Zapier} & 5 & 51 & 4 & 4 & 40 &
\href{https://evernote.com/compare-plans}{Evernote} & 6 & 32 & 4 & 0 & 4 \\ \hline
\rowcolor{gray!20}
\href{https://slack.com/pricing}{Slack} & 4 & 44 & 4 & 4 & 21 &
\href{https://hypercontext.com/pricing}{Hypercontext} & 4 & 63 & 4 & 0 & 4 \\ \hline
\href{https://mailchimp.com/es/pricing/marketing/compare-plans/?currency=USD}{MailChimp} & 6 & 90 & 4 & 5 & 15 &
\href{https://pumble.com/pricing}{Pumble} & 4 & 34 & 4 & 0 & 4 \\ \hline
\rowcolor{gray!20}
\href{https://clickup.com/pricing}{ClickUp} & 6 & 135 & 4 & 2 & 13 &
\href{https://userguiding.com/pricing}{UserGuiding} & 5 & 59 & 3 & 1 & 4 \\ \hline
\href{https://planable.io/pricing/}{Planable} & 6 & 41 & 4 & 2 & 13 &
\href{https://www.crowdcast.io/pricing}{Crowdcast} & 5 & 16 & 3 & 0 & 3 \\ \hline
\rowcolor{gray!20}
\href{https://clockify.me/pricing}{Clockify} & 6 & 72 & 6 & 4 & 10 &
\href{https://www.deskera.com/pricing}{Deskera} & 4 & 100 & 3 & 0 & 3 \\ \hline

\href{https://www.microsoft.com/en-us/microsoft-365/enterprise/office365-plans-and-pricing}{Microsoft 365} & 6 & 60 & 4 & 1 & 8 &
\href{https://www.overleaf.com/user/subscription/plans}{Overleaf} & 6 & 16 & 3 & 0 & 3 \\ \hline
\rowcolor{gray!20}
\href{https://trustmary.com/pricing/}{Trustmary} & 5 & 45 & 4 & 1 & 8 &
\href{https://quip.com/about/pricing}{Quip} & 6 & 15 & 3 & 0 & 3 \\ \hline
\end{tabularx}
\caption{SaaS dataset with the number of snapshots and some metrics related to their 2024 pricing. S indicates the number of available snapshots; F the number of features; P the number of plans; A the number of add-ons; and C the size of the configuration space.} 
\label{table:dataset}
\end{table}

\textbf{Pricing modeling}. Given that, to the best of our knowledge, Pricing4SaaS (see Fig. \ref{fig:pricing4saas}) is the only metamodel available in the literature that represents SaaS pricings, we decided to model all the studied pricings using its YAML serialization: Yaml4SaaS. This approach not only validate the feasibility of this pricing model but also provides a dataset with enough portability for further research in Pricing-driven SaaS DevOps. In addition, to minimize human error and mismatches during the modeling phase, several rules were established:

\begin{enumerate}
    
    \item Features containing ``and'' in their description are separated into distinct features, e.g. ``Record and play audio notes'' of \href{https://web.archive.org/web/20221130112813/https://evernote.com/compare-plans}{Evernote 2022} must be splitted into ``Record audio notes'' and ``play audio notes'' features. 
    
    \item Plans and add-ons without a clear feature list are not modeled, e.g. ``Teams'' plan of \href{https://web.archive.org/web/20221130112813/https://evernote.com/compare-plans}{Evernote 2022}. 
    
    
    \item Features offered as a limited trial or demo for a specific plan, thus not providing any permanent access, will not be included within the feature set of such plan. E.g. \href{https://web.archive.org/web/20231130184417/https://slack.com/pricing}{Slack's 2023} ``Free'' plan does not include ``file history''.

    \item Recommended user limits must not be not modeled, as they do not restrict any other feature.
\end{enumerate}


\subsection{Saas pricings analysis results}
Table \ref{table:dataset} provides an overview of the metrics computed for the models of the 30 SaaS for 2024. 
Next, we report the results with regard to our research questions on SaaS pricings' configuration and evaluation spaces evolution.

\begin{figure}[htbp]
    \centering
    \begin{subfigure}{0.49\textwidth}
        \centering
        \includegraphics[width=\textwidth]{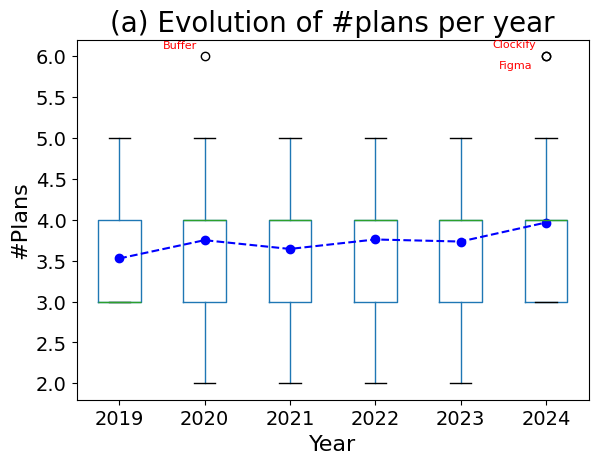}
        \phantomsubcaption
        \label{fig:subfig1}
    \end{subfigure}
    \hfill
    \begin{subfigure}{0.49\textwidth}
        \centering
        \includegraphics[width=\textwidth]{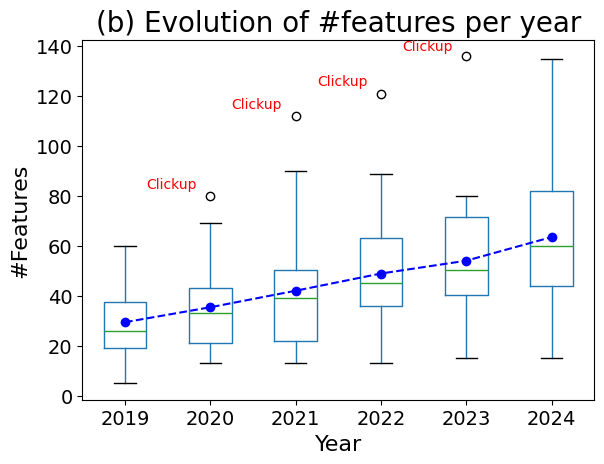}
        \phantomsubcaption
        \label{fig:subfig2}
    \end{subfigure}
    \vspace{-0.08\textwidth} 

    \begin{subfigure}{0.49\textwidth}
        \centering
        \includegraphics[width=\textwidth]{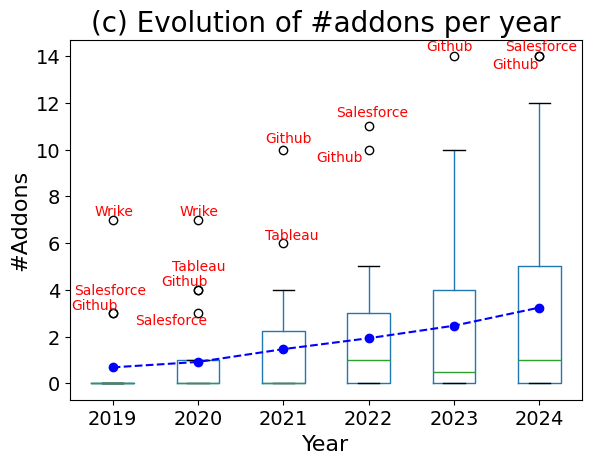}
        \phantomsubcaption
        \label{fig:subfig3}
    \end{subfigure}
    \begin{subfigure}{0.49\textwidth}
        \centering
        \includegraphics[width=\textwidth]{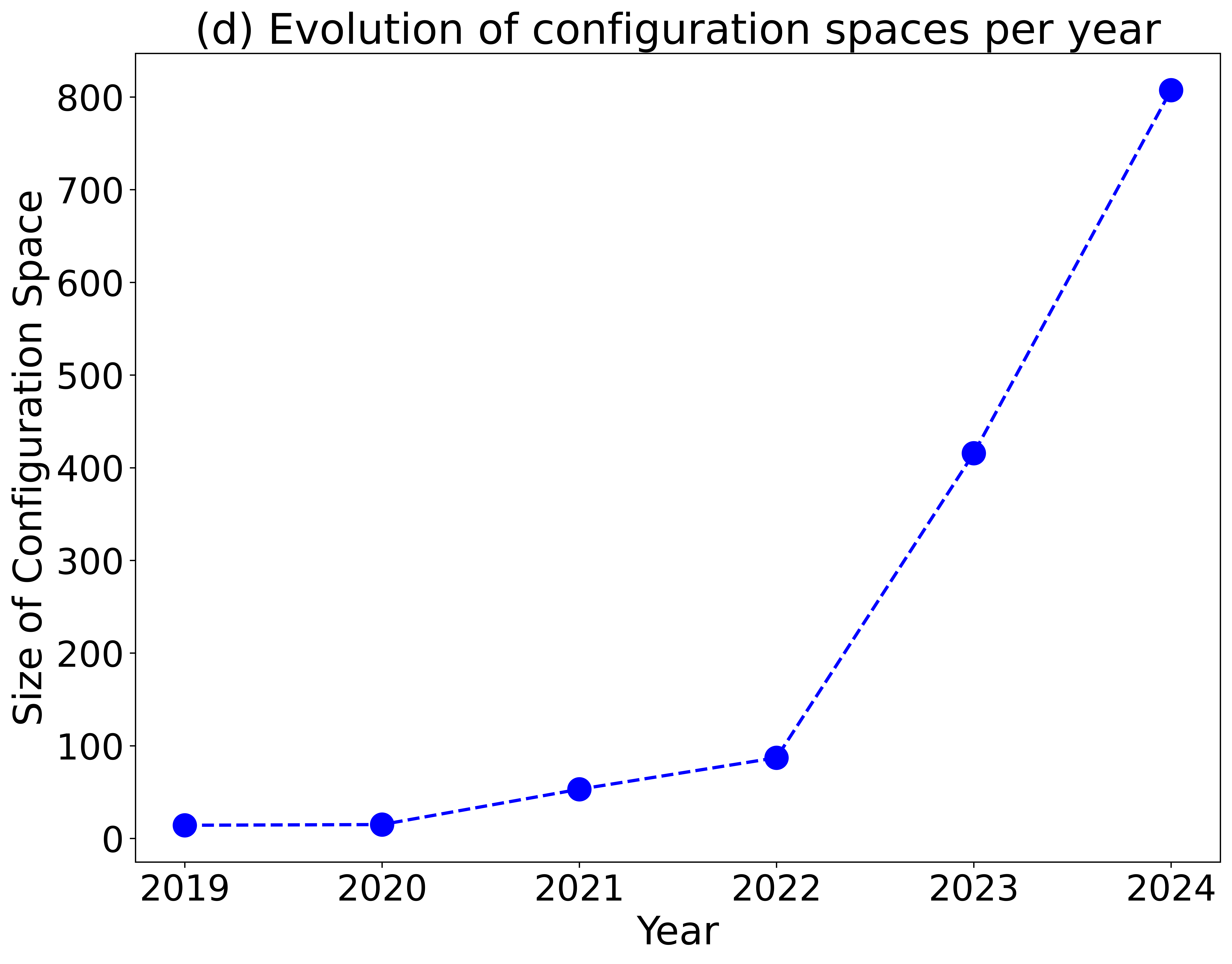}
        \phantomsubcaption
        \label{fig:subfig4}
    \end{subfigure}
    \vspace{-0.06\textwidth}
    \caption{Evolution of SaaS pricings per year.
    }
    \label{fig:wholefigure}
\end{figure}

%
Fig. \ref{fig:wholefigure} depicts the trends observed in the aspects addressed by $RQ_{1}$ across the sampled SaaS pricings. The box plots are enriched with a dashed blue line that connects the mean values for the whole set of SaaS per year, which helps to visualize the evolution of the indicator more clearly. This is in addition to displaying the median (indicated by the green line within the box) and the overall distribution of values.


\emph{Features growing rate}. The results (Fig. \ref{fig:subfig2}) show a linear increase in the number of features over the years. From 2019 to 2024, the average number of features of the SaaS under study has increased by 115\% (from 29.47 to 63.40). 
\emph{Evolution of the numbers of plans.} The number of plans has shown a relatively stable trend from 2019 to 2024 (Fig. \ref{fig:subfig1}), with the number of plans hovering around three to four (the mean increase between 2019 and 2024 is an 11\%). While there are occasional outliers, such as \textit{Buffer} and \textit{Clockify}, the overall distribution does not show significant increases or decreases in the number of plans offered. This suggests that while some SaaS providers may experiment with the number of plans, the general approach across the industry has remained consistent in terms of the number of pricing plans available to customers. The reason for this tendency might be the Paradox of Choice (see Section \ref{sec:background:saasPricing}), as keeping a reduced number of plans facilitates the decision of the customer. 


\emph{Evolution of the number of add-ons}. As shown in Fig. \ref{fig:subfig3}, there is a distinct trend towards incorporating more add-ons within SaaS pricing models, with projections indicating a linear increase in the future. Between 2019 and 2024, the average number of add-ons has increased by a 363\% (from 0.68 to 3.16). This trend may be attributed to the fact that add-ons expand the configuration space of a pricing model without falling into the Paradox of Choice, as discussed in Section \ref{sec:background:saasPricing}.
Fig. \ref{fig:subfig4} shows an exponential increase of the cofiguration space due to the increase in the number of add-ons, validating empirically our formulation of the size of the configuration space (see Section \ref{sec:background:saasPricing}). 

\emph{Statistical tests}, specifically the Mann-Whitney U tests, were conducted based on the results of normality and homoscedasticity tests, to compare the number of plans, features, and add-ons between 2019 and 2024. The differences were statistically significant across all compared magnitudes: plans, features, and add-ons. Effect sizes, calculated using the r estimator, indicate medium to large effects for the differences in the number of features and large effects for the differences in the number of plans and add-ons. This means the statistically significant differences identified also have a large practical impact. Detailed results and specific p-values are available in the supplementary laboratory package associated with this paper \cite{LABPACK}. These results support our general research question $RQ_1$, showing a vibrant pace of change and rapid evolution towards increased complexity in both the SaaS pricings and the configuration and evaluation spaces.

\begin{center}
    \scalebox{0.95}{
\begin{tcolorbox}[top=2pt,bottom=2pt,left=2pt,right=2pt,title=Answers to \textbf{RQ\textsubscript{1}: How are pricings’ configuration and evaluation spaces evolving in modern SaaS?}]
    \begin{rqAnswer}
        \small
            \item Configuration spaces are experimenting an exponential rise due to the increase in the number of add-ons.
            \item The increasing number of features is expanding the size of evaluation spaces, making the management of SaaS pricings more challenging.
        \normalsize
    \end{rqAnswer}
\end{tcolorbox}
}
\end{center}

\subsection {Methodology for the analysis of feature toggling support}
\label{sec:FeatureTogglingStudyMethodology}
\textbf{Comparison levels}. Since the purpose of this study is not to perform a general comparative of the features provided by the feature toggling solutions, but to evaluate their suitability and support to streamline pricing-driven feature toggling, the capability levels presented in Section \ref{sec:featureToggles} are used.

\textbf{Sample selection}. In order to choose the specific set of tools to be compared in this study, we applied a methodology that combines keyword search with snowballing. First, we performed a search on Google using the keywords ``feature toggling tool'' and ``feature toggling library'' as search terms, resulting in the identification of five tools. Next, we read the description and documentation of each identified tool, looking for links to other feature toggling solutions. Finally, as a second form of snowballing, we performed a search on \url{alternativeto.net} using the name of each tool already identified as the search term, leading to the identification of the remaining tools in the sample.

As a result of applying this methodology, the set of tools identified and the results of the evaluation of the capabilities are shown in Table \ref{tab:feature-toggling-evaluation}. 

\begin{table}[htbp]
    \centering
    \renewcommand{\arraystretch}{1.1}
    \setlength{\tabcolsep}{4pt}
    \begin{tabular}{|l|c|c|c|l|c|c|c|}
        \hline
        \textbf{Product / Company} & \textbf{L1} & \textbf{L2} & \textbf{L3} & \textbf{Product / Company} & \textbf{L1} & \textbf{L2} & \textbf{L3} \\
        \hline
        \rowcolor{gray!20}  
            \href{https://www.abtasty.com/}{Abtasty} & \checkmark & \checkmark & & 
            \href{https://apptimize.com/}{Apptimize} & \checkmark & & 
            \\            
            \href{https://configcat.com/}{ConfigCat} & \checkmark & \checkmark &  & 
            \href{https://devcycle.com/}{DevCycle} & \checkmark & \checkmark & $\sim$ 
            \\                            
        \rowcolor{gray!20}  
            \href{https://engineering.fb.com/2017/08/31/web/rapid-release-at-massive-scale/}{Facebook's Gatekeeper} & \checkmark & & & 
            \href{https://www.flagsmith.com/}{FlagSmith} & \checkmark & \checkmark & 
            \\
            \href{https://www.featurehub.io/}{FeatureHub} & \checkmark & \checkmark & & 
            \href{https://www.harness.io/}{Harness} & \checkmark & &  
            \\
        \rowcolor{gray!20}                              
            \href{https://github.com/jayf/javascript-feature-flags}{javascript-feature-flags} & & & & 
            \href{https://launchdarkly.com/}{LaunchDarkly} & \checkmark & \checkmark & $\sim$
            \\
            \href{https://www.molasses.app/}{Molasses} & \checkmark & & &
            \href{https://openfeature.dev/}{OpenFeature} & \checkmark & \checkmark &
            \\
        \rowcolor{gray!20}
            \href{https://www.optimizely.com}{Optimizely} & \checkmark & \checkmark & & 
            \href{https://pricing4saas-docs.vercel.app}{Pricing4Saas} & \checkmark & \checkmark & \checkmark  \\
            \href{https://github.com/paralleldrive/react-feature-toggles/}{react-feature-toggles} & & & & \href{https://www.statsig.com/}{StatSig} & \checkmark & \checkmark &  \\                            
        \rowcolor{gray!20}  
            \href{https://tggl.io/}{Tggl} & \checkmark & \checkmark & &
            \href{https://www.togglz.org/}{Togglz} & \checkmark & \checkmark & $\sim$             
             \\
             \href{https://tweek.soluto.io/}{Tweek} & \checkmark & \checkmark & &
            \href{https://github.com/Unleash/unleash}{Unleash} & \checkmark & \checkmark & $\sim$              
            \\
            \rowcolor{gray!20}  
            \href{https://engineering.linkedin.com/ab-testing/xlnt-platform-driving-ab-testing-linkedin}{XLNT} & \checkmark & \checkmark & & & &  & \\
        \hline
    \end{tabular}
    \caption{Evaluation of various products/companies against levels L1, L2, and L3. A checkmark (\checkmark) indicates that the product/company achieves the capability level, and a $\sim$ indicates that the level is achieved partially.}
    \label{tab:feature-toggling-evaluation}
    \vspace{-1.0cm}
\end{table}

\subsection{Feature toggling solutions analysis results}
Next, we report the results with regard to our research questions on the capabilities of feature toggling solutions to streamline and automate pricing-driven feature toggling.
Regarding, \emph{configuration-based toggling} (level \emph{L1}), the majority of feature toggling tools support this capability (90.4\%), as it is also used for purposes such as user segmentation. 
However, the \emph{dynamic and extensible contextual evaluation (L2)} is not as widely supported, with a 71.4\% of the tools (15 out of 21) providing this feature. Finally, the \emph{pricing-aware evaluation (L3)}, is only fully supported by our suite of tools: Pricing4SaaS \cite{ICWE24}, but 4 out the 21 industrial and open-source solutions provide partial support (19\%). This means that those tools include the extension mechanisms required to implement pricing awareness. 


\begin{center}
    \scalebox{0.95}{
    \begin{tcolorbox}[top=2pt,bottom=2pt,left=2pt,right=2pt,title=Answers to \textbf{RQ\textsubscript{2}: Are current feature toggling tools appropriate for optimizing pricing-driven feature toggling?}]
        \begin{rqAnswer}
            \small
                \item Only one of the feature toggling solutions currently supports the implementation of pricing-driven feature toggling.
                \item A small set of industrial solutions 
                has the potential to support pricing-driven feature toggling, as these included solutions possess the necessary elements.
            \normalsize
        \end{rqAnswer}
    \end{tcolorbox}
    }
\end{center}

\subsection{Limitations identified in Yaml4SaaS}


\begin{center}
    \scalebox{0.95}{
    \begin{tcolorbox}[top=2pt,bottom=2pt,left=2pt,right=2pt,title=Answers to \textbf{RQ\textsubscript{3}: Does Yaml4SaaS provide enough expressiveness to represent real world pricings?}]
        \begin{rqAnswer}
            \small
                \item We were able to model most of the complexities of the pricings in our sample.
                \item The model does not support defining formulas for computing numeric values, such as the price of plans/add-ons or an usage limit. E.g. if a plan's price is calculated using a mathematical expression, such as ``tasks'' in \href{https://web.archive.org/web/20240709201542/https://zapier.com/pricing}{Zapier 2024}.
                \item Yaml4SaaS is limited in that it cannot model inter-add-on dependencies or other complex restrictions. E.g. \href{https://web.archive.org/web/20221119101619/https://www.microsoft.com/en-us/security/business/endpoint-security/microsoft-defender-business}{Microsoft Defender 2022} highlights the need for add-ons that depend on other add-ons. 
                \item Yaml4SaaS cannot model custom subscription periods, such as semesters, it only supports monthly and annual periods. A new approach to support the definition of such periods is needed.
            \normalsize
        \end{rqAnswer}
    \end{tcolorbox}
    }    
\end{center}

\section{Threats to validity}
\label{sec:threadsToValidity}

The factors that could have influenced our study and how they were mitigated are summarized in the following threats to internal and external validity.

\textbf{Internal validity}. Internal threats to validity encompass various factors that can introduce bias or distort the data without our awereness, resulting in inaccurate conclusions \cite{wohlin2012}. The main source of bias is the subjective and manual process applied for modeling the SaaS pricings using the Yaml4SaaS syntax, leading to missed or misclassified features. This introduces potential \textit{measurement bias}. To mitigate this threat, we used as our baseline, the pricings previously modeled from the study that introduced Yaml4SaaS syntax \cite{FIRST_EXPERIMENT}, ensuring a consistent and reliable foundation. Each SaaS pricing model included data spanning at least four years, which helps identify and correct anomalies over time since features typically persist between consecutive years. Moreover, multiple authors contributed to the modeling process and regular cross-verification sessions were conducted to ensure consistency and accuracy along the process, addressing \textit{observer bias}. Additionally, following the specific methodology developed for this task (see Section \ref{sec:MethodologyForTheAnalysisOfPricingStructures}), helped to minimize individual biases and errors. The impact of possible mistakes was also minimized by the large number of SaaS modeled: 30 unique SaaS for a six years gap (although not every SaaS has a model for every year), making a total of 162 SaaS pricings modeled, which makes us remain confident of the overall accuracy of the results, mitigating \textit{selection bias}. 

\textbf{External validity}. Threats to external validity relate to the degree to which we can generalize from the analysis \cite{wohlin2012}. Our study addresses \textit{population validity} examining a substantial but limited number of SaaS (30) and feature toggling tools (21). To minimize this threat, we systematically selected a broad range of real-world SaaS from multiple domains and popular feature toggling tools, including some with millions of users worldwide. We employed techniques such as keyword searches in public repositories and snowballing (see Sections \ref{sec:MethodologyForTheAnalysisOfPricingStructures} and \ref{sec:FeatureTogglingStudyMethodology}) to ensure \textit{cultural validity} by including a diverse and representative sample. 

\section{Related Work}
\label{sec:relatedWork}

\agarcia{Hay que reestructurar esta sección en base al comentario de Antonio, con especial interés en la relacion del espacio de configuración presentado por galán}

\aruiz{es habitual expresar el RW en el mismo orden en el que s ehan presentado conceptos. P.e., en nuestro caso hablar primero del pricing y después del feature toggling. El pricing compararlo con el pricing de APIs y el de IaaS, ... el concepto de espacio de configuración de iaas .. Cuando se hable del IaaS, podemos enlazar la aproximación de galán como una en la que el espacio de configuración del pricing también representa .... y que se automatiza su análisis con FMs, el modelo pro excelencia en el ámbito de las SPL para respresentar de manera compacta el espacio de configuració, y que nos hemos inspirado en el cross--tree constraint 'requires' y 'excludes' para la definición que hemos propuesto de espacio de configuración.   comentar tb la relación entre los usage limits y los límites de capacidad de APIs }
The concept of pricing-driven feature toggling represents an evolution in the management and dynamic adaptation of multi-tenant single-instance SaaS applications. Previous studies, such as \cite{Fatma2017}, explored the variability in functional and non-functional requirements for different tenants within the same application instance. For example, their work detailed scenarios where two tenants could request the same feature/service but with different data schemes to meet specific security, availability, and performance needs. They employed a Model-Driven Engineering (MDE) approach combined with Software Product Line (SPL) techniques, requiring a feature model to manage this variability.

In contrast, the pricing-driven feature toggling approach diverges by utilizing the application's pricing model as its foundation rather than a traditional feature model. This method enables the direct translation of pricing constraints into feature toggles, streamlining the adaptation process to cater to varied tenant requirements without the need for a dedicated service version for each customer.

Our study identifies a gap in standardized pricing models for SaaS, similar to the issues highlighted by \cite{gamez2017analysis} in the context of RESTful APIs in 2017. In their analysis, they pointed out the ad-hoc and platform-dependent nature of pricing plan modeling in APIs, emphasizing the lack of standardization that hindered the creation of API governance tools at that time. They aimed to facilitate standardization and modeling of usage limits for APIs through a detailed industry analysis. Drawing a parallel, our research addresses a similar need in the SaaS pricing domain, advocating for standardized pricing models to enhance Pricing-driven Development and Operation of SaaS. 


The increasing complexity of SaaS pricing models, as evidenced by our analysis in the evolution of pricings' configuration and evaluation spaces, highlights the need for efficient tools and frameworks that manage such complexity. Although most of the current industry feature toggling tools support dynamic toggling and extensibility, their support for pricing plan sensitivity remains limited, with only an academic tool \cite{ICWE24} currently addressing this need. This gap presents an opportunity for further development to fully realize the potential of pricing-driven feature toggling in SaaS.


\section{Conclusions and Future Work}
\label{sec:conclusions}

In this paper, we have focused on the possibilities of current technology to reduce the time required to transfer changes in SaaS pricing to the software and underlying infrastructure, without compromising its quality and reliability.

We have found that the trend in pricing is to increase the configuration space by increasing the number of add-ons. More specifically, the size of the configuration space, i.e., the number of different configurations of plans and add-ons offered by a pricing, grows exponentially with the number of add-ons and linearly with the number of plans. If we see the configuration space as an iceberg, the plans and add-ons are just the tip of the iceberg. Furthermore, we found that the evaluation space is also growing, given the sustained increase in pricing features, which is greater than 100\% in six years.

We have also found that feature toggling, specifically permission toggles, appears to be a promising technique for allowing SaaS software to adapt to pricing changes by simply making changes to the YAML file that serializes the pricing. This technique is promising for two main reasons: first, although we have only found one library that achieves this, its approach can serve as inspiration for replication in different technologies using the numerous feature toggling tools available; and second, because Yaml4SaaS has been shown to have sufficient expressiveness to model pricings of real-world SaaS.

In conclusion, this study provides solid evidence that motivates the development of automated solutions, or those with minimal human intervention, to reduce the time-to-market of SaaS updates that result from changes in pricings, i.e., feature configurations and their usage limits. Furthermore, the technological development necessary to adapt current feature toggling tools to realize the development of pricing-driven SaaS has been identified and is feasible, especially if a standard for serializing pricing is adopted.

Several challenges remain for future work, but we focus primarily on modeling more real pricings, developing useful solutions by improving current libraries, and validating them in real-world scenarios. 

 \vspace{-0.2cm}
 \section*{Replicability \& verifiability}
 \vspace{-0.1cm}
 All the artifacts and datasets generated in this study are available in the laboratory package of the study \cite{LABPACK}. This material comprises of the companion technical report, our serialization of each pricing
 , and the raw data-set and jupyter notebook used to generate the figures and statistical results reported.

\bibliographystyle{splncs04}
\bibliography{references}

\begin{thebibliography}{10}
\providecommand{\url}[1]{\texttt{#1}}
\providecommand{\urlprefix}{URL }
\providecommand{\doi}[1]{https://doi.org/#1}

\bibitem{LABPACK}
Anonymous: {SaaS Analysis} - supplementary material (2024), \url{https://anonymous.4open.science/r/icsoc24-saas-analysis/}

\bibitem{saasSource1}
Cruz, L.: 37 saas examples you need to know about in 2024 (2024), \url{https://clickup.com/blog/saas-examples/}, accessed: June 2024

\bibitem{FOWLER2023}
Fowler, M.: Feature toggles (aka feature flags) (nd), \url{https://martinfowler.com/articles/feature-toggles.html}, accessed: December 2023

\bibitem{gamez2017analysis}
Gamez-Diaz, A., Fernandez, P., Ruiz-Cortés, A.: {An analysis of RESTful APIs offerings in the industry}. In: International Conference on Service-Oriented Computing. pp. 589--604. Springer (2017)

\bibitem{JCIS2024LONG}
Garc\'{i}a-Fern\'{a}ndez, A., Parejo, J.A., Ruiz-Cort\'{e}s, A.: {Pricing-driven Development and Operation of SaaS: Challenges and Opportunities}. In: {Actas de las XIX Jornadas de Ciencia e Ingenier\'{i}a de Servicios (JCIS)}. SISTEDES (2024)

\bibitem{FIRST_EXPERIMENT}
Garc{\'i}a-Fern{\'a}ndez, A., Parejo, J.A., Ruiz-Cort{\'e}s, A.: {Pricing4SaaS} - supplementary material (2023), \url{https://doi.org/10.5281/zenodo.10292553}

\bibitem{CAISEFORUM24}
Garc{\'\i}a-Fern{\'a}ndez, A., Parejo, J.A., Ruiz-Cort{\'e}s, A.: Pricing4saas: Towards a pricing model to drive the operation of saas. In: International Conference on Advanced Information Systems Engineering, (CAiSE). pp. 47--54. Springer (2024)

\bibitem{ICWE24}
Garc{\'i}a-Fern{\'a}ndez, A., Parejo, J.A., Trinidad, P., Ruiz-Cort{\'e}s, A.: {Towards Pricing4SaaS: A Framework for Pricing-Driven Feature Toggling in SaaS}. In: Web Engineering. ICWE. pp. 389--392. Springer (2024)

\bibitem{GHADDAR2012}
Ghaddar, A., Tamzalit, D., Assaf, A., Bitar, A.: Variability as a service: Outsourcing variability management in multi-tenant saas applications. In: Proceeding of the International Conference of Advanced Information Systems Engineering, (CAiSE). pp. 175--189 (2012)

\bibitem{Jiang2009}
Jiang, Z., Sun, W., Tang, K., Snowdon, J., Zhang, X.: A pattern-based design approach for subscription management of software as a service. 2009 Congress on Services - I pp. 678--685 (2009)

\bibitem{Fatma2017}
Mohamed, F., Mizouni, R., Abu-Matar, M., Al-Qutayri, M., Whittle, J.: An integrated platform for dynamic adaptation of multi-tenant single instance saas applications. In: Proceedings - 2017 IEEE 5th International Conference on Future Internet of Things and Cloud. pp. 257--264 (2017)

\bibitem{Rahman2016}
Rahman, M.T., Querel, L.P., Rigby, P.C., Adams, B.: Feature toggles: Practitioner practices and a case study. Proceedings - 13th Working Conference on Mining Software Repositories, MSR 2016 pp. 201--211 (5 2016)

\bibitem{schwartz2015paradox}
Schwartz, B.: The paradox of choice. Positive psychology in practice: Promoting human flourishing in work, health, education, and everyday life pp. 121--138 (2015)

\bibitem{TERNAVA2022}
Tërnava, X., Lesoil, L., Randrianaina, G.A., Khelladi, D.E., Acher, M.: On the interaction of feature toggles. In: Proceedings of the 16th International Working Conference on Variability Modelling of Software-Intensive Systems (2022)

\bibitem{wohlin2012}
Wohlin, C., Runeson, P., Höst, M., Ohlsson, M.C., Regnell, B., Wesslén, A.: Experimentation in Software Engineering. Springer Berlin, Heidelberg (2012)

\end{thebibliography}

\end{document}